\documentclass[12pt]{article}

\usepackage{bm}
\usepackage[margin=25mm]{geometry}

\usepackage{appendix}
\usepackage{hyperref}
\usepackage{xcolor}
\usepackage{amsmath}

\usepackage{graphicx}
\usepackage{authblk}

\providecommand{\keywords}[1]
{
  \small	
  \textbf{\textit{Keywords---}} #1
}

\title{Estimating causal effects in the presence of competing events using regression standardisation with the Stata command standsurv}

\author[1]{Elisavet Syriopoulou}
\author[2]{Sarwar I Mozumder}
\author[2]{Mark J Rutherford}
\author[1,2]{Paul C Lambert}

\affil[1]{Department of Medical Epidemiology and Biostatistics, Karolinska Institutet, Stockholm, Sweden}
\affil[2]{Biostatistics Research Group, Department of Health Sciences, University of Leicester, Leicester, United Kingdom}

\date{}

\begin{document}

\maketitle

\begin{abstract}
When interested in a time-to-event outcome, competing events that prevent the occurrence of the event of interest may be present.
In the presence of competing events, various statistical estimands have been suggested for defining the causal effect of treatment on the event of interest.
Depending on the estimand, the competing events are either accommodated or eliminated, resulting in causal effects with different interpretation.
The former approach captures the total effect of treatment on the event of interest while the latter approach captures the direct effect of treatment on the event of interest that is not mediated by the competing event.
Separable effects have also been defined for settings where the treatment effect can be partitioned into its effect on the event of interest and its effect on the competing event through different causal pathways.
We outline various causal effects that may be of interest in the presence of competing events, including total, direct and separable effects, and describe how to obtain estimates using regression standardisation with the Stata command \texttt{standsurv}. 
Regression standardisation is applied by obtaining the average of individual estimates across all individuals in a study population after fitting a survival model.
With \texttt{standsurv} several contrasts of interest can be calculated including differences, ratios and other user-defined functions.
Confidence intervals can also be obtained using the delta method.
Throughout we use an example analysing a publicly available dataset on prostate cancer to allow the reader to replicate the analysis and further explore the different effects of interest.

\end{abstract}


\section{Background}

When a time-to-event outcome is of interest, other events may preclude the event of interest, which means that it cannot be observed.
For instance, when investigating survival in a population with prostate cancer, the event of interest is often death due to prostate cancer.
However, some individuals might die due to other causes and therefore the occurrence of a death to prostate cancer is not observed. 
These types of events are known as competing events \cite{Putter2007, Andersen2012}. 
For simplicity, in this paper we focus on time-to-death outcomes, however the methods are applicable to any time-to-event outcome (e.g. time to relapse). Currently, there is a growing interest in the estimation of causal effects for treatment in the presence of competing events for the event of interest; these are contrasts under different treatment arms that have a causal interpretation given some assumptions \cite{Hernan2004a}.  
Defining the causal effect in a competing events setting can be complex and requires special consideration on dealing with the competing events. 

Various estimands of interest have been suggested where the competing events are either accommodated or eliminated, with each approach requiring a different set of assumptions and having a different interpretation. 
The former approach is conducted using crude measures e.g. cause-specific cumulative incidence functions (also known as crude probabilities of death in population-based cancer research) that is
the risk an individual dies from the cause of interest \cite{Kalbfleisch2002}. 
If a patient is at high risk of dying from a competing cause, then this will reduce their risk of dying from the cause of interest.
Crude measures are useful for patients and clinicians as they quantify risk in a real-world setting and they can also aid in policy decisions e.g. on resource allocation \cite{Eloranta2013, Belot2019}. 
In the latter approach of eliminating competing risks, net measures, such as net probability of death, are estimated instead. 
The net probability of death corresponds to a hypothetical world  where competing events cannot occur.
Net measures are useful for comparing survival between different populations such as countries or socio-economic groups as they are not affected by other-cause mortality \cite{Lambert2015}. 
They can also be of great interest for studying the aetiology of a disease or temporal trends \cite{Morris2016, Eloranta2013a}. 

Even though several statistical estimands have been suggested before in the presence of competing events, these are often described without the use a formal causal framework making interpretation of the estimating effects cumbersome \cite{Geskus2016, Austin2016, Latouche2013}.
Recent work by Young \textit{et al.} \cite{Young2020} utilised a counterfactual framework to explicitly describe each of the classical statistical estimands and define causal effect as well as their identifying assumptions when competing events exist.
Based on whether the competing events are defined as censoring events or not, the authors defined contrasts of risk as either the direct effect of the treatment on the event of the interest that is not mediated by the competing event or the total effect of treatment on the event of interest. 
In settings in which the treatment exerts its effect on the event of interest and its effect on the competing event through different causal pathways, so called separable effects have been defined \cite{Stensrud2020}.
The separable direct effect is the treatment effect on the event of interest that is not mediated by its effect on the competing event. 
The separable indirect effect is the treatment effect on the event of interest that is only through its effect on the competing event. 

Causal effects are identifiable under certain assumptions and can be estimated using regression standardisation or inverse probability weighting \cite{Hernan2006, Young2020}.
Doubly robust approaches such as doubly robust standardisation have also been suggested \cite{Vansteelandt2011a, Funk2011}.
In this paper, we focus on regression standardisation methods.
To estimate the average causal effect with regression standardisation, first a survival model is fitted and then predictions are obtained for every individual in the study population under each fixed treatment arm \cite{Sjolander2016}. 
An average of the individual-specific estimates is calculated, and the relevant contrasts between treatment arms (such as the difference between treatment arms) are formed.
Regression standardisation has recently been utilised for obtaining estimates of various estimands in the presence of competing events. 
Mozumder \textit{et al.} \cite{Mozumder2021} applied regression standardisation for estimating the restricted mean failure time, which is the average life-years lost before a pre-specified time in the presence of competing events, after fitting a single Royston-Parmar flexible parametric model on either the log-cumulative subdistribution or cause-specific hazards scale. 
The authors also partitioned the total number of years lost into the number of years lost due to each cause of death. 
Kipourou \textit{et al.} \cite{Kipourou2019} estimated cause‐specific cumulative probabilities using flexible regression models for the cause‐specific hazards and applying regression standardisation for marginal estimates.

In this paper, we outline direct and total effects as well as separable effects that may be of interest in the presence of competing events and describe how to obtain estimates of those using regression standardisation with the Stata command \texttt{standsurv}. 
Throughout we use an example utilising a publicly available dataset on prostate cancer to allow the reader to replicate the analysis and further explore the measures.
Stata code for all the analysis is also available at \url{https://github.com/syriop-elisa/competing_events_standsurv}.

The paper is structured as follows.
In Section \ref{sec:illustrativeExample} we introduce the illustrative example and describe the models we will use for the analyses.
In section \ref{sec:nocomp} we briefly introduce statistical estimands and define the causal effect in the absence of competing events.
Next, in Section \ref{sec:comp}, we define causal effects in the presence of competing events: total, direct and separable effects, and show how to obtain those using regression standardisation with command \texttt{standsurv}.
A discussion of the methods is provided in Section \ref{sec:discussion}.

\section{Introducing the illustrative example} \label{sec:illustrativeExample}
\begin{figure}[h!]
    \centering
    \includegraphics[width=0.9\textwidth]{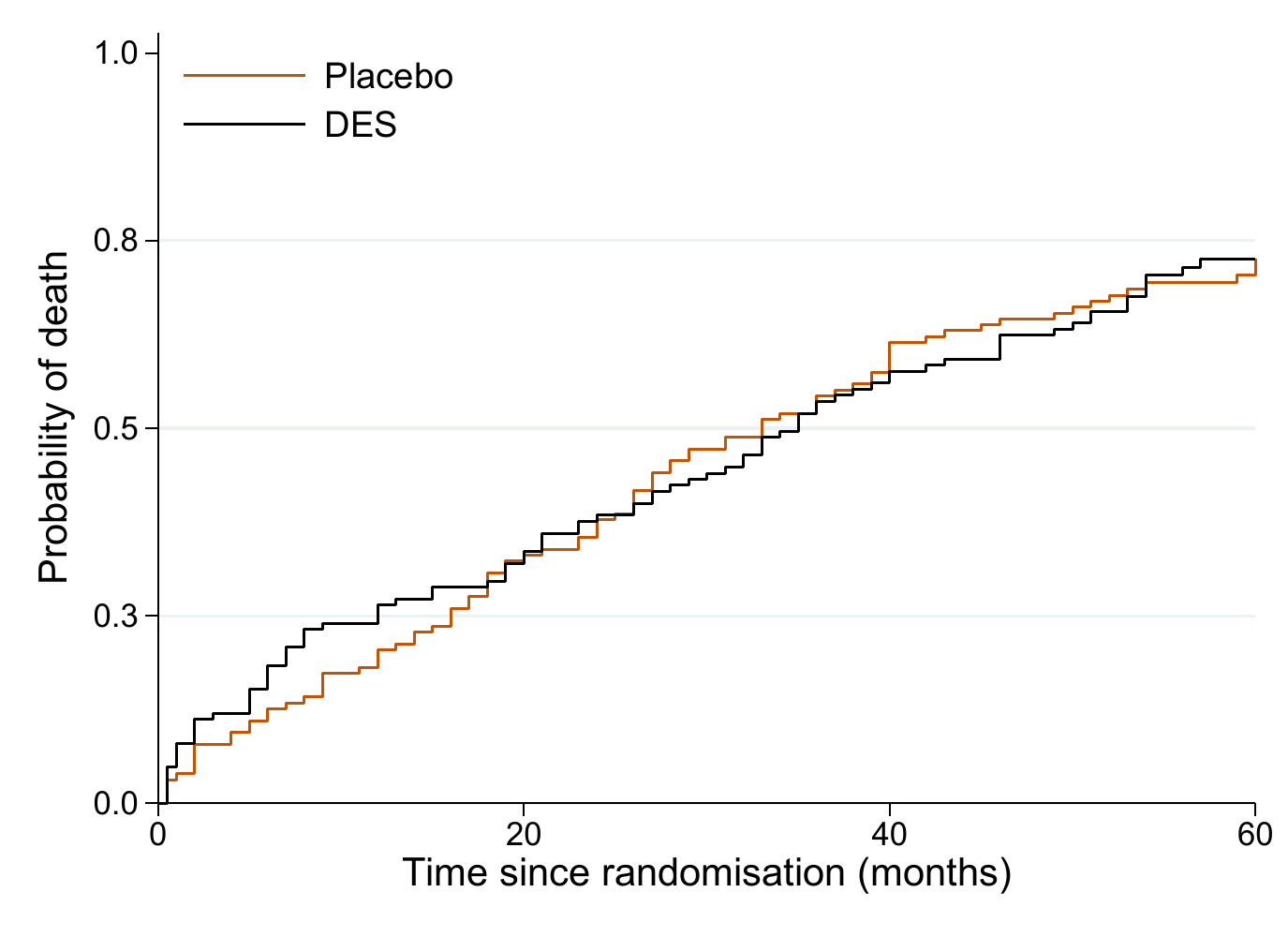}
    \caption{Kaplan-Meier failure curves for all-cause deaths by treatment group.}
    \label{fig:KMfailure}
\end{figure}

For the remainder of the paper we use data from a trial on prostate cancer (\texttt{prostate.dta}) to demonstrate how to obtain several measures of interest using Stata.
This dataset has been used in several methodological papers, including the recent papers by Young \textit{et al.} \cite{Young2020} and Stensrud \textit{et al.} \cite{Stensrud2020}.
Data include 502 individuals that were randomly assigned estrogen therapy and are available at \url{https://hbiostat.org/data} \cite{Byar1980}.
There are four treatment arms but for simplicity we restrict our analysis to high-dose estrogen therapy arm (i.e. diethylstilbestrol, DES) and placebo.
We are interested on the causal effect of treatment on prostate cancer death and death due to other causes is considered a competing event.
For simplicity, we categorised all continuous variables and code for this can be found in the Appendix \ref{sec:appendix}.
We chose the same cut-offs as in the Young \textit{et al.} \cite{Young2020} paper, while Stensrud \textit{et al.} \cite{Stensrud2020} chose slightly different cut-offs.
For the analysis, we will use user-written Stata commands; a list of these and with information on how to install the commands in Stata is also available in Appendix \ref{sec:appendix}. 
The following variables will be used in our analysis:
\texttt{rx}: treatment arm (1: DES, 0: placebo),
\texttt{hgBinary}: hemoglobin level (1: $<$12 (g/100ml), 0: $\geq$12),
\texttt{ageCat}: age (0: 0-59, 1: 60-74, 2: 75-100 years),
\texttt{hx}: history  of cardiovascular disease (with values 0 and 1),
\texttt{normalAct}: daily activity function (1: normal activity, 0: otherwise)
\texttt{dtime}: months of follow-up,
\texttt{eventType}: cause of death (0: alive, 1: dead due to prostate cancer, 2: dead due to other causes).
The Kaplan-Meier failure curves for all-cause deaths by treatment group is shown in Figure \ref{fig:KMfailure}.
The first months after randomisation the DES group has a higher probability of death from any cause in comparison to the placebo group.
However, approximately 20 months after randomisation the curves cross for the first time and remain close to each other up to 60 months, suggesting that treatment has almost a negligible effect on all-cause death probability.

To explore the data further, we will need to fit two cause-specific models; one for the event of interest i.e. prostate cancer death and one for the competing event i.e. all other causes of death.
First we need to declare the data as survival data.
To declare the survival data with the event being defined as death due to other causes, \texttt{eventType==2}, the following command can be used:
\begin{verbatim}
stset dtime, failure(eventType==2) exit(time 60)
\end{verbatim}

The \texttt{exit} option restricts follow-up time to 60 months (5 years) since randomisation and censor those still alive after that or those with prostate cancer deaths.

The analysis of the data will be performed using flexible parametric survival models (FPMs), so called Royston-Parmar models. 
Flexible parametric survival modelling is a methodology that was first introduced by Royston and Parmar and allows a wide range of hazard functions by using restricted cubic splines for the effect of time \cite{Royston2011}. 
FPMs have many advantages in terms of modelling time-dependent effects and making predictions.
Flexible parametric models can be fitted within Stata using the user-written command \texttt{stpm2}.
Factor variables are not supported and dummy variables must be generated before fitting the models.
For instance, to fit a FPM survival model in the log cumulative hazard scale (option \texttt{scale(hazard)}) including treatment, daily activity function, age, history of cardiovascular diseases and hemoglobin level, assuming 3 degrees of freedom (that is equal to the number of knots used to create the splines minus 1) for the baseline hazard: 
\begin{verbatim}
// First, create dummy variables (ageCat1 - ageCat3) for age
tab ageCat, gen(ageCat)
// Then, fit the model using option eform to obtain hazard ratios
stpm2 rx normalAct ageCat2 ageCat3 hx hgBinary, scale(hazard) df(3) eform
\end{verbatim}
The above model gives the following output:
\begin{verbatim}
Log likelihood =  -297.4793                     Number of obs     =        252

------------------------------------------------------------------------------
             |     exp(b)   Std. Err.      z    P>|z|     [95% Conf. Interval]
-------------+----------------------------------------------------------------
xb           |
          rx |   1.313613    .243568     1.47   0.141     .9133536    1.889278
   normalAct |   .9325932   .2690789    -0.24   0.809     .5297818    1.641676
     ageCat2 |   2.695251   1.596876     1.67   0.094     .8438812    8.608293
     ageCat3 |   3.573784   2.141445     2.13   0.034     1.104282    11.56582
          hx |   2.241948   .4361735     4.15   0.000     1.531169    3.282674
    hgBinary |   1.858763   .4248224     2.71   0.007     1.187629    2.909156
       _rcs1 |   2.895811   .2676688    11.50   0.000     2.415967    3.470958
       _rcs2 |   .9164784   .0494298    -1.62   0.106     .8245427    1.018665
       _rcs3 |    .944936   .0324615    -1.65   0.099     .8834073     1.01075
       _cons |     .06403   .0421594    -4.17   0.000     .0176167    .2327242
------------------------------------------------------------------------------
\end{verbatim}

Here, \texttt{\_rcs1} -- \texttt{\_rcs3} are the splines used to model the baseline hazard. 
In this model, the youngest age group (\texttt{ageCat1}) is omitted from the model and is used as the reference.
The model assumes proportional hazards and thus the hazard ratio (HR) for DES compared to placebo remains constant across follow-up with a HR of 1.31 which is suggestive of a harmful effect of the treatment for other cause mortality.

We can also store the model estimates as \texttt{other} to use them later
\begin{verbatim}
  estimates store other  
\end{verbatim}

Similarly to the model fitted above for deaths from other causes, we fit a cause specific model for death due to prostate cancer.
Once again we need to \texttt{stset} the data and we define the event of interest as \texttt{eventType==1}.
\begin{verbatim}
stset dtime, failure(eventType==1) exit(time 60)
\end{verbatim}

We then fit a FPM model for death due to prostate cancer and this time we assume that there are time dependent effects:

\begin{verbatim}
stpm2 rx normalAct ageCat2 ageCat3 hx hgBinary, scale(hazard) df(4) eform ///
     tvc(rx) dftvc(2)
\end{verbatim}
Time-dependent effects are allowed in the model using the option \texttt{tvc()} to indicate the variables (in this example treatment) and \texttt{dftvc()} to denote the number of degrees of freedom for the time-dependent effects.
We obtain the following output:

\begin{verbatim}
Log likelihood = -175.95468                     Number of obs     =        252

------------------------------------------------------------------------------
             |     exp(b)   Std. Err.      z    P>|z|     [95% Conf. Interval]
-------------+----------------------------------------------------------------
xb           |
          rx |    .781722   .2573405    -0.75   0.454     .4100523    1.490271
   normalAct |   .3363049   .1239687    -2.96   0.003     .1632914    .6926327
     ageCat2 |   .5847662   .2359638    -1.33   0.184     .2651594    1.289607
     ageCat3 |   .8689135   .3752287    -0.33   0.745     .3727319    2.025613
          hx |   .5876147   .1792134    -1.74   0.081     .3232134    1.068307
    hgBinary |   1.615195   .5107219     1.52   0.129     .8691131    3.001742
       _rcs1 |    4.65067   1.473811     4.85   0.000     2.499005    8.654936
       _rcs2 |   1.163117   .2192974     0.80   0.423      .803777    1.683106
       _rcs3 |   .9559556   .0545688    -0.79   0.430     .8547687    1.069121
       _rcs4 |   1.061586   .0356486     1.78   0.075     .9939658    1.133807
    _rcs_rx1 |   .8307078   .3447018    -0.45   0.655     .3683347    1.873501
    _rcs_rx2 |   .7526368   .1743185    -1.23   0.220     .4780115    1.185039
       _cons |   .6245197   .3314558    -0.89   0.375     .2206896      1.7673
------------------------------------------------------------------------------
\end{verbatim}
Terms \texttt{\_rcs\_rx1} -- \texttt{\_rcs\_rx2} correspond to an interaction between treatment and time (time-dependent effect).
In the above model, treatment is allowed to a have a time-dependent effect and the HR is changing over time so it can not be obtained directly from the output. 
For instance, when comparing hazard rates for an individual in the DES group and an individual in the placebo group, with both individuals belonging in the same groups of all adjusting covariates, the HR is 0.52 at 12 months since randomisation, 0.9 at 36 months since randomisation and 1.5 at 60 months since randomisation which is suggestive of a protective effect of the treatment in the short-term for prostate cancer mortality.

We will also store the model estimates as \texttt{prostate} to use it later on.
\begin{verbatim}
estimates store prostate
\end{verbatim}

The cause-specific models described above are simplified models with no interactions that we will consider for the remaining sections to demonstrate how to obtain causal effects using the postestimation command \texttt{standsurv}.
Interactions and non-linear effects can also be modelled and these are discussed in Appendix \ref{sec:advanceModelling}.
As mentioned earlier, for the applied example we fit FPMs on the log cumulative hazard scale.
However, \texttt{standsurv} also supports FPMs on log hazard scale as well as standard parametric models.

We also create a variable for the time points at which we want to obtain predictions. 
Below we create a variable called \texttt{timevar} that includes 121 timepoints from time 0 to 60 months (every half month):
\begin{verbatim}
range timevar 0 60 121
\end{verbatim}


\section{When no competing events exist} \label{sec:nocomp}
Let $X$ denote treatment and let also $\bm{Z}$ denote a set of measured confounders that is sufficient for confounding control.
Lowercase letters, such as $x$, denote a specific (fixed) level for treatment while lowercase letters with subscript $i$, such as $\bm{z_i}$ denote the observed value of an individual $i$.
Let the conditional probability of the event of interest at time $t$ be $F(t| X,\bm{Z})$.
Assume that the event of interest is death due to any cause so that there are no competing events and assume also non-informative censoring.
The marginal counterfactual all-cause probability of death had all individuals in the population, possibly contrary to fact, been assigned $X=x$ is 

\begin{equation}
    \label{noCE}
E[F(t| X=x,\bm{Z})]
\end{equation} 
with the expectation taken over the marginal distribution of $\bm{Z}$ and $F(t)=1-S(t)$ with $S(t)$ denoting the all-cause survival. 
Equation \ref{noCE} is conceptually similar to equation 1 in Young \textit{et al.} that define the estimand of interest in the absence of competing events as the counterfactual risk of the event.
The average causal difference can be defined as
\begin{equation}
    E[F(t| X=1,\bm{Z})]-E[F(t| X=0,\bm{Z})]  
\end{equation}
with the first term being the all-cause probability of death when setting $X=1$ and the second term is the all-cause probability of death when setting $X=0$ for everyone in the study population.

Under assumptions, the marginal probability of death defined in Equation \eqref{noCE} can be estimated by the standardised probability of death using regression standardisation.
After fitting a survival model, individual-specific predictions are obtained for everyone in the study population (of size $N$) and these are averaged over the marginal distribution of the observed covariate pattern $\bm{Z }=\bm{z_i}$:
\begin{equation}\label{eq:total-ac:est}
   E[\widehat{F}(t| X=x,\bm{Z})] =  \frac{1}{N} \sum_{i=1}^{N}\widehat{F}(t| X=x,\bm{Z}=\bm{z_i})]
\end{equation}
Estimates for all other estimands described below will also be obtained using regression standardisation with the command \texttt{standsurv}, as the average of individual-specific predictions as described in Equation \eqref{eq:total-ac:est}.
More information on regression standardisation can be found elsewhere \cite{Syriopoulou2020, Sjolander2016}.

\section{When competing events exist} \label{sec:comp}
Often, competing events that prevent the occurrence of the event of interest will be present.
For instance, in our illustrative example where the event of interest is death due to prostate cancer, death due to other causes act as a competing event.
In the presence of competing events the cause-specific hazard functions are defined as
\begin{equation}
    h_k(t)= \lim_{\Delta t \to 0} \frac{P[t\leq T<t+\Delta t, D=k|T\geq t]}{\Delta t}
\end{equation}
with $D$ denoting the cause of death i.e. $k=c$ if the event of interest is death due to prostate cancer and $k=o$ for death due to other causes.

The cause-specific survival functions can also be defined through the standard transformation from hazard to survival function; let $S_c(t)$ and $S_o(t)$ denote the prostate cancer and other cause survival respectively. 

In the presence of competing events there are several statistical estimands that may be of interest depending on the research question.
In Section \ref{sec:total} we define total effects of treatment that refer to a real-world setting where both competing events are present (accommodating competing events) and in Section \ref{sec:direct} we define direct effects in a hypothetical world where the only possible cause of death is prostate cancer (eliminating competing events).

\subsection{Total effects}\label{sec:total}
Various estimands can be defined for the total treatment effect on survival.
Such measures accommodate competing events and entail no hypothetical interventions regarding censoring of competing events.
These are often referred to as crude measures.
Crude measures are highly relevant for patients and health professionals.
Patients and clinicians want to know the actual survival in a real-world setting where other causes of death are present.
In addition to being more relevant for clinicians and healthcare professionals, crude measures can also aid in policy decisions e.g. on resource allocation.


\subsubsection{Cause-specific cumulative incidence functions}\label{sec:total:crudeprob}
The total effect of treatment on the event of interest can be defined using cause-specific cumulative incidence functions (CIFs).
Let $F_c (t|X=x,\bm{Z})$ denote the counterfactual cumulative incidence for death due to prostate cancer when setting treatment to $X=x$.
The marginal counterfactual cumulative incidence of prostate cancer death in the presence of death due to other causes as the competing event is defined as:

\begin{equation}\label{crudeprobcancer}
    E\left[F_c (t|X=x,\bm{Z})\right]= E\left[\int_{0}^{t}S(u|X=x,\bm{Z})h_c(u|X=x,\bm{Z})du\right]
\end{equation} 
where $S(u|X=x,\bm{Z})=S_c(u|X=x,\bm{Z})S_o(u|X=x,\bm{Z})$ is the all-cause survival and $h_c(u|X=x,\bm{Z})$ is the prostate cancer hazard.
The average causal difference is defined as
\begin{equation}\label{crudeprobcancerdif}
    E\left[F_c (t|X=1,\bm{Z})\right] - E\left[F_c (t|X=0,\bm{Z})\right]
\end{equation}
and refers to the total effect of treatment (through all causal pathways) on prostate cancer death and includes those possibly mediated by the competing event.

Similarly, the marginal counterfactual cumulative incidence of death due to other causes in the presence of death due to prostate cancer as the competing event is:
\begin{equation}\label{crudeprobotherdif}
    E\left[F_o (t|X=1,\bm{Z})\right] - E\left[F_o (t|X=0,\bm{Z})\right]
\end{equation}
with 
\begin{equation}\label{crudeother}
  E\left[F_o (t|X=x,\bm{Z})\right]= E\left[\int_{0}^{t}S(u|X=x,\bm{Z})h_o(u|X=x,\bm{Z})du\right]
\end{equation}
where $h_o(u|X=x,\bm{Z})$ is the cause-specific hazard for other causes.
Equations \ref{crudeprobcancer} and \ref{crudeother} are conceptually similar to equations 7 and 9 in the paper by Young \textit{et al.} \cite{Young2020}.

\emph{Example}\\[2ex]
The cause-specific cumulative incidence functions can be estimated by fitting separate models for each cause of death.
Recall that we have already \texttt{stset} the prostate cancer data and have fitted cause-specific flexible parametric survival models for death due to prostate cancer and death due to other causes, in Section \ref{sec:illustrativeExample}.

By applying regression standardisation with command \texttt{standsurv} we can obtain the standardised cause-specific CIFs under DES and under placebo.
For this, we use option \texttt{cif}  that requests the cause-specific CIFs to be estimated (the default is overall survival) and the option \texttt{crmodels(cancer other)} that gives the names of the cause-specific model estimates.  
Each of the model estimates need to have been stored in memory previously using \texttt{estimates store}.
\begin{verbatim}
standsurv, crmodels(prostate other) cif at1(rx 0) at2(rx 1) timevar(timevar) ///
           contrast(difference) ci atvars(CIF0 CIF1) contrastvars(CIF_diff)          
\end{verbatim}

Each of the \texttt{atn()} options creates a standardised CIF based on the fixed covariate values specified in the \texttt{atn()} options.
Above, with the \texttt{at1()} option we force the covariate \texttt{rx} to be set to 0 (placebo) for all subjects and then in the \texttt{at2()} option we force the covariate \texttt{rx} to be set to 1 (DES) for all subjects. 
The key point is that the distribution of the remaining confounders is forced to be the same under DES and placebo and any covariates not specified in the \texttt{atn()} options keep their observed values. 
For instance, if instead of the observed age distribution, we wanted to obtain predictions as every patients belonged to the oldest age group (ageCat3), then this could be obtained by setting age group \texttt{ageCat2} to take the value 0 and age group \texttt{ageCat3} to take the value 1 within the \texttt{atn()} options.
The \texttt{contrast()} option asks for a comparison of the two CIFs (under DES and under placebo) with the \texttt{difference} argument asking to take differences in the CIFs. 
By default \texttt{at1()} is the reference, i.e. the contrast will be \texttt{at2}--\texttt{at1}, but this can be changed using the \texttt{atref()} option.
Option \texttt{atvar()} gives the names of the new variables to be created for each \texttt{atn()} option and \texttt{contrastvar()} gives the new variables to be created when using the \texttt{contrast()} option.
In the example above, the following variables are created: the CIFs of death due to prostate cancer will be  \texttt{CIF0\_prostate} under placebo, \texttt{CIF1\_prostate} under DES, \texttt{CIF\_diff\_prostate} for their difference, and similarly the CIFs of death due to other causes will be \texttt{CIF0\_other} under placebo, \texttt{CIF1\_other} under DES, \texttt{CIF\_diff\_other} for their difference. 
As the \texttt{ci} option was specified there will be upper and lower bounds for the confidence interval (CI, 95\% by default) for each estimate.
Standard errors for the estimates are obtained using the delta method \cite{Cox2005, Lambert2009}.

\begin{figure}[h!]
    \centering
    \includegraphics[width=0.92\textwidth]{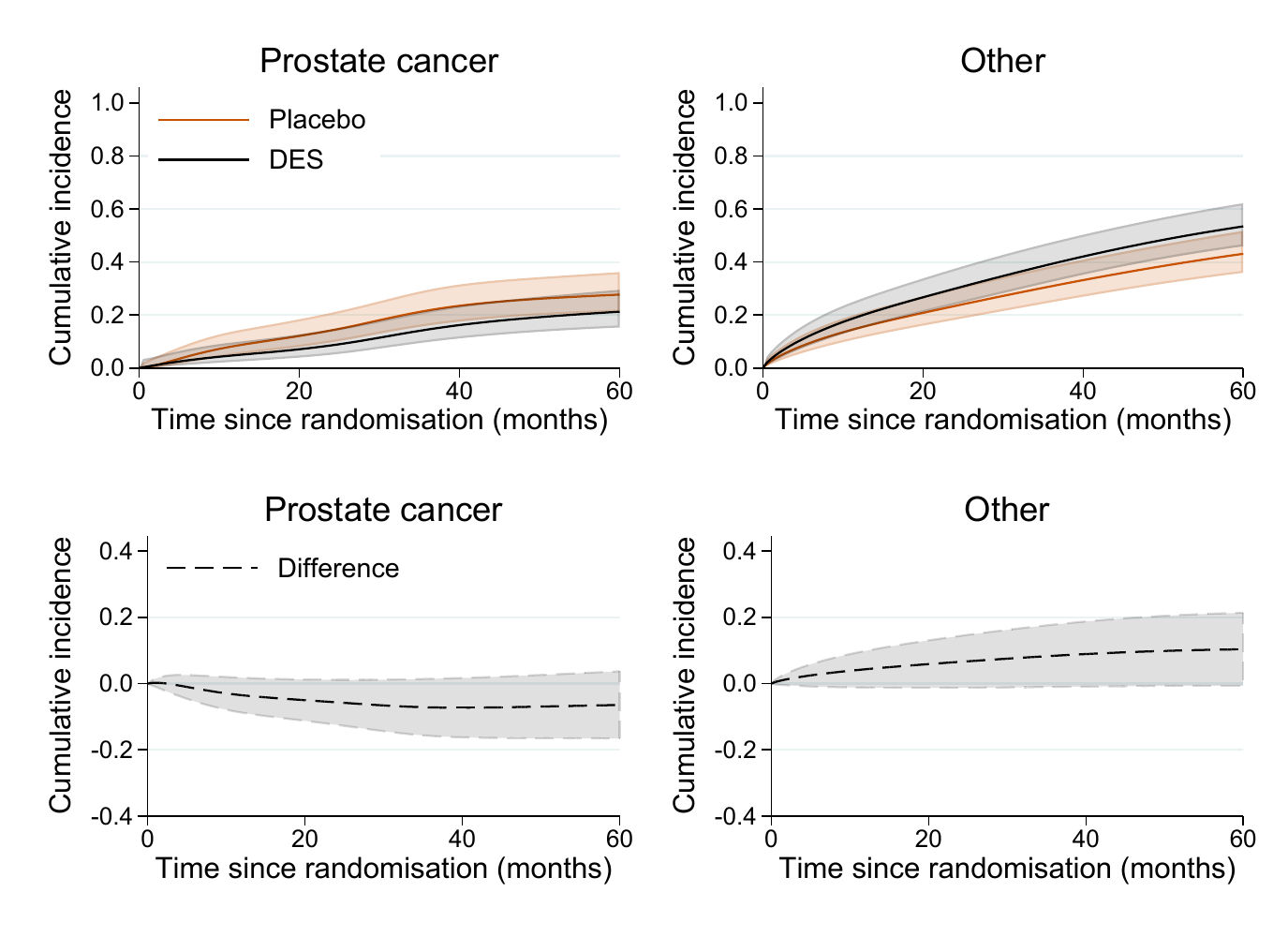}
    \caption{Standardised cumulative incidence of prostate cancer death and other cause of death under DES and under placebo and the difference between treatment arms, with 95\% confidence intervals.}
    \label{fig:total:crude}
\end{figure}

Figure \ref{fig:total:crude} shows the standardised CIFs of prostate cancer death and CIFs of other cause of death under DES, under placebo as well as their difference by time since randomisation.
Sixty months (5 years) after randomization, the CIF of prostate cancer death is equal to 21.3\% (95\% CI: 15.3\%-29.5\%) under DES while under placebo is higher and equal to 27.7\% (95\% CI: 21.2\%-36.2\%).
For the CIFs of death from other cause the pattern is reversed and is much higher; in particular it is equal to 53.5\% (95\% CI: 45.9\%-62.2\%) under DES and equal to 43.1\% (95\% CI: 35.9\%-51.7\%) under placebo.
The above estimates account for the fact that competing events are also present. 

\subsubsection{Expected loss in life due to a cause of death}
Another estimand for the total effect of treatment is the expected life lost before time $t^*$ \cite{Andersen2013, Mozumder2021}.
The marginal counterfactual expected life lost before time $t^*$ (also referred to as restricted mean failure time (RMFT)) when setting treatment to $X=x$ is defined as
\begin{equation}
    L(0, t^*|X=x, \bm{Z}) = E\left[\sum_{k=1}^{K} \int_{0}^{t^*}F_k(u|X=x, \bm{Z})du\right]
\end{equation}

with $k$ denoting the cause of death.
The RMFT can also  be partitioned further to the life lost due to each cause $k$ before time $t^*$

\begin{equation}
    L_k(0, t^*|X=x, \bm{Z}) = E\left[\int_{0}^{t^*}F_k(u|X=x,\bm{Z})du\right]
\end{equation}
The marginal counterfactual difference in expected loss in life due to cause $k$ before time $t^*$ when setting $X=1$ and setting $X=0$ can then be defined as:

\begin{equation}
    L_k(0, t^*|X=1, \bm{Z})-  L_k(0, t^*|X=0, \bm{Z})
\end{equation}

The expected life lost corresponds to a comparison of the study population to an immortal cohort where all individuals remain alive at the end of the follow-up period at time  $t^*$.
Even though the expected years lost is a useful measure for exploring the impact of different causes, this comparison can make interpretation of the measure challenging as the comparison involves a hypothetical construct.

\emph{Example}\\[2ex]
For the estimation of the expected loss in life in the prostate cancer data example, we need to select a timepoint $t^*$. 
The estimates will vary by the choice of $t^*$.
Here we choose 60 months:
\begin{verbatim} 
gen t_rmft60 = 60 in 1
\end{verbatim}
After fitting cause-specific models, the standardised expected loss in life due to each cause can be obtained by using the option \texttt{rmft} together with options \texttt{crmodels()} and \texttt{cif}:
\begin{verbatim}
standsurv, crmodels(prostate other) cif rmft  ///
    at1(rx 0) at2(rx 1) timevar(t_rmft60) contrast(difference) ci ///
    atvars(RMFT0 RMFT1) contrastvars(RMFT_diff)
\end{verbatim}

To list the estimates of the life lost due to prostate cancer before 60 months:
\begin{verbatim}
// under placebo
list t_rmft60 RMFT0_prostate* in 1, noobs abb(22)       
+---------------------------------------------------------------------+
| t_rmft60   RMFT0_prostate   RMFT0_prostate_lci   RMFT0_prostate_uci |
|---------------------------------------------------------------------|
|       60        10.112996            7.5136987            13.611498 |
+---------------------------------------------------------------------+

//under DES
list t_rmft60 RMFT1_prostate* in 1, noobs abb(22)
+---------------------------------------------------------------------+
| t_rmft60   RMFT1_prostate   RMFT1_prostate_lci   RMFT1_prostate_uci |
|---------------------------------------------------------------------|
|       60        6.9136108            4.7205369            10.125546 |
+---------------------------------------------------------------------+

// their difference
list t_rmft60 RMFT_diff_prostate* in 1, noobs abb(22)
+---------------------------------------------------------------------------------+
| t_rmft60   RMFT_diff_prostate   RMFT_diff_prostate_lci   RMFT_diff_prostate_uci |
|---------------------------------------------------------------------------------|
|       60           -3.1993855               -7.2043426                .80557166 |
+---------------------------------------------------------------------------------+
\end{verbatim}

Similarly, to list the estimates of the life lost due to other causes before 60 months:
\begin{verbatim}
// under placebo
list t_rmft60 RMFT0_other* in 1, noobs abb(22)
+------------------------------------------------------------+
| t_rmft60   RMFT0_other   RMFT0_other_lci   RMFT0_other_uci |
|------------------------------------------------------------|
|       60     15.637513         12.644666         19.338733 |
+------------------------------------------------------------+

// under DES    
list t_rmft60 RMFT1_other* in 1, noobs abb(22) 
+------------------------------------------------------------+
| t_rmft60   RMFT1_other   RMFT1_other_lci   RMFT1_other_uci |
|------------------------------------------------------------|
|       60     19.813057         16.498763         23.793132 |
+------------------------------------------------------------+

// their difference     
list t_rmft60 RMFT_diff_other* in 1, noobs abb(22)
+------------------------------------------------------------------------+
| t_rmft60   RMFT_diff_other   RMFT_diff_other_lci   RMFT_diff_other_uci |
|------------------------------------------------------------------------|
|       60         4.1755443            -.54768438             8.8987729 |
+------------------------------------------------------------------------+
\end{verbatim}

When 60 months were chosen, death due to other causes resulted in more months lost than prostate cancer.
The number of months lost due to other causes was also higher under DES; 19.8 (95\%: 16.5 --23.8)) months under DES in comparison to 15.6 (95\%:12.6--19.3) months under placebo, resulting in a difference of 4.2 (95\%: -0.6 -- 8.9) months.
The number of months lost due to prostate cancer was, however, higher under placebo;  under DES 6.9 (95\%: 4.7- 10.1) months were lost while under placebo 10.1 (95\%: 7.5- 13.6) months were lost, resulting in a difference of  -3.2 (95\%: -7.2 -- 0.8) months between DES and placebo.

We can also calculate the total expected loss in life as the sum of the months lost from each cause and it quantifies the average months of life that a patient lost from time 0 up to a pre-defined timepoint $t^*$ \cite{Calkins2018, Royston2011b, Royston2013, Chen2001}. 
Even though this can also be obtained after fitting an all-cause model, here we show how to obtain estimates after fitting cause-specific models.
The total number of months lost due to all causes can be obtained within \texttt{standsurv} using option \texttt{lincom(\#...\#)} that calculates a linear combination of \texttt{atn()} options and it also provides confidence intervals using the delta method.
Option \texttt{lincom(\#...\#)} is used here instead of the \texttt{contrast()} option that we used above to calculate the difference between \texttt{atn()} options.
For the total months lost under placebo, the first two \texttt{\#} in \texttt{lincom()} that correspond to \texttt{at1()} should be set to 1 (these refer to the months lost due to prostate cancer and the months lost due to other causes): 
\begin{verbatim}
standsurv, crmodels(prostate other) cif rmft  ///
    at1(rx 0) at2(rx 1) timevar(t_rmft60) lincom(1 1 0 0) ci ///
    atvar(RMLT0b RMLT1b) lincomvar(RMLT_total0)
\end{verbatim}
The total months lost within 60 months since randomisation under placebo were 25.8 (95\%: 22.3--29.3) months:
\begin{verbatim}
li RMLT_total0* in 1, noobs abb(22)
+-------------------------------------------------+
| RMLT_total0   RMLT_total0_lci   RMLT_total0_uci |
|-------------------------------------------------|
|   25.750509         22.255255         29.245764 |
+-------------------------------------------------+
\end{verbatim}

Similarly, the total months lost under DES are obtained by setting the last two \texttt{\#} in lincom() to 1 (these refer to the months lost due to prostate cancer and the months lost due to other causes):
\begin{verbatim}
standsurv, crmodels(prostate other) cif rmft  ///
    at1(rx 0) at2(rx 1) timevar(t_rmft60) lincom(0 0 1 1) ci ///
    atvar(RMFT0c RMFTc) lincomvar(RMFT_total1)
\end{verbatim}
The total months lost within 60 months since randomisation under DES were 26.7 (95\%: 23.2 --30.2) months:
\begin{verbatim}
li RMFT_total1* in 1, noobs abb(22)
+-------------------------------------------------+
| RMFT_total1   RMFT_total1_lci   RMFT_total1_uci |
|-------------------------------------------------|
|   26.726668         23.223267          30.23007 |
+-------------------------------------------------+
\end{verbatim}
This results in a difference of total loss of approximately 1 month  between placebo and DES and is effectively the same as the sum of the differences calculated above for each specific cause (-3.2 and 4.2).

\subsection{Direct effects}\label{sec:direct}
Consider a hypothetical intervention that sets  $S_o(t|X = x, Z) = 1$ i.e an intervention that eliminates the competing deaths due to other causes.
Contrasts of counterfactuals between different level of the treatment under such intervention are controlled direct effects which quantify the treatment's effect on the event of interest not mediated by competing events.

Let $F_c^{N}(t)$ denote the net probability of death due to prostate cancer.
The marginal counterfactual probability of death under an intervention of eliminating competing events when setting $X=x$ is given by
\begin{equation}
    E\left[F_c^{N} (t|X=x,\bm{Z})\right]= E\left[\int_{0}^{t}S_c(u|X=x,\bm{Z})h_c(u|X=x,\bm{Z})du\right]
\end{equation}

This is similar to Equation \eqref{crudeprobcancer}, but here $S_o(t|X = x, Z)$ is omitted from the integral.

The average causal difference in net probabilities of prostate cancer death if competing events were eliminated is then defined as:
\begin{equation}
    \label{dif_csa}
    E[F_c^{N}(t| X=1, Z)] - E[F_c^{N}(t| X=0, Z)]
\end{equation}

The above equations are conceptually similar to equations 5 and 6 in the paper by Young et al  \cite{Young2020}.
Despite their interpretation in a hypothetical world, net measures are useful for comparing different populations such as countries or socioeconomic groups as they are not affected by other-cause mortality (i.e. mortality due to other causes) \cite{Eloranta2020}.
They can also be of great interest for studying the aetiology of a disease or temporal trends.

\emph{Example}\\[2ex]
The net probability of prostate cancer death under DES and placebo as well as their difference, if competing events were eliminated, can be obtained by applying regression standardisation as follows.
For this only the cause-specific model for prostate cancer death will be considered (the estimates of this model were stored earlier under \texttt{prostate}).
All other competing events are censored.

\begin{figure}[h!]
    \centering
    \includegraphics[width=0.9\textwidth]{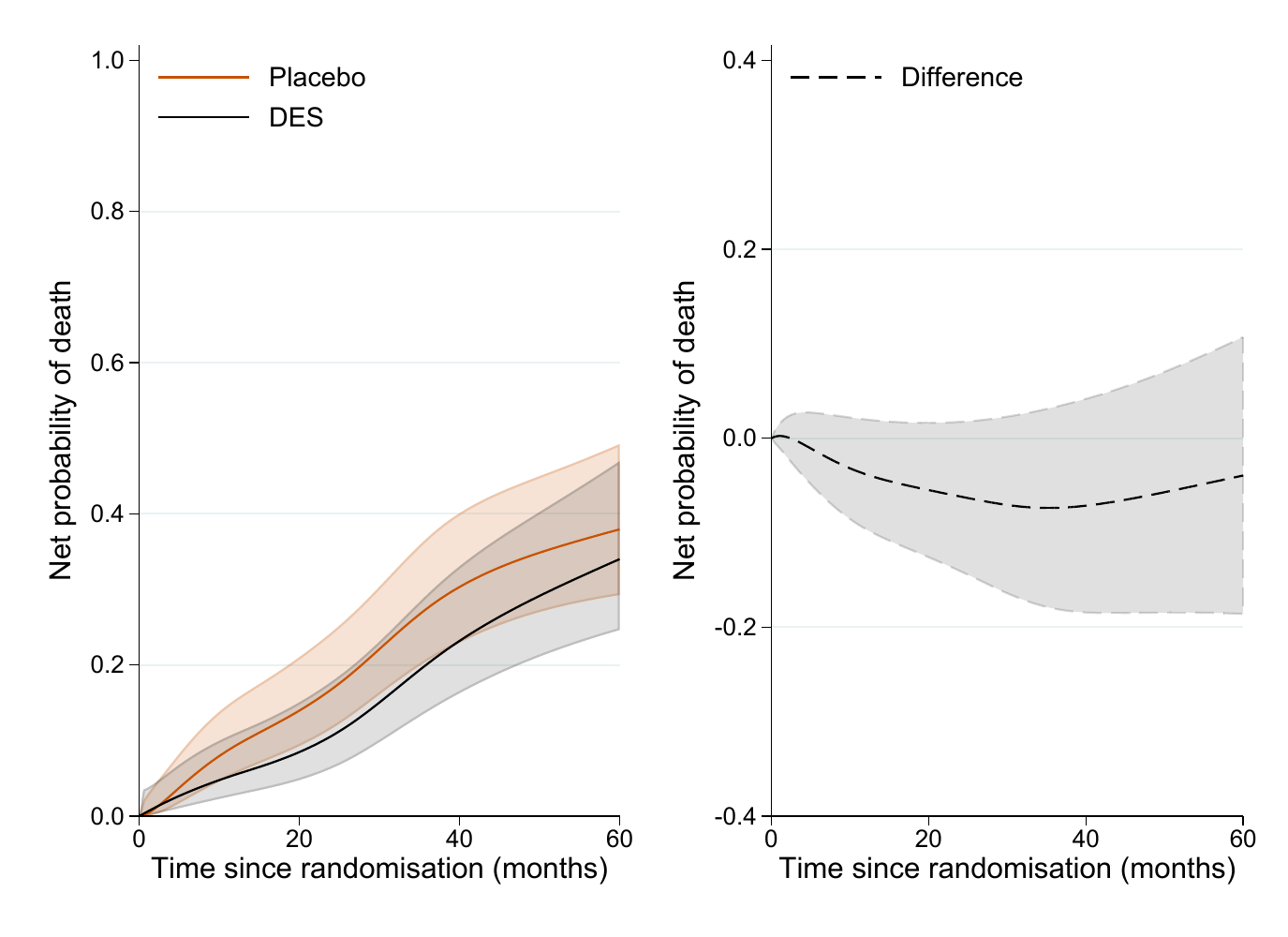}
    \caption{Standardised net probability of death from prostate cancer under DES and under placebo and their difference by time since randomisation, with 95\% confidence intervals.}
    \label{fig:direct:netsurv}
\end{figure}

We load the model estimates under \texttt{prostate} and use the post estimation command \texttt{standsurv} with option \texttt{failure}:
\begin{verbatim}
estimates restore prostate

standsurv, failure at1(rx 0) at2(rx 1) timevar(timevar) contrast(difference) ci /// 
    atvars(F_net_prostate0 F_net_prostate1) contrastvars(F_net_prostate_diff)
\end{verbatim}

Figure \ref{fig:direct:netsurv} shows the standardised net probability of death under DES and placebo as well as their difference by time since randomisation.   
Sixty months  after  randomisation,  the standardised  net  probability of death from prostate cancer under DES  was  equal  to  34\% (95\% CI: 24.6\%--47\%) and under placebo 38\% (95\% CI: 29.2\%--49.2\%) ,  resulting  in  a  difference  of  -4\% (95\%: -18.6\%--10.7\%).  
In contrast to Section \ref{sec:total}, these estimates assume that prostate cancer is the only possible cause of death and that it is not possible to die from other causes.
Such interpretation might be challenging, however it allows to capture the direct effect of treatment on prostate cancer mortality that is not mediated by competing events. 

\subsection{Separable effects}\label{sec:separable}
In situations where the treatment effect can be decomposed into distinct components, separable effects can be estimated \cite{Stensrud2020}.
Suppose that the treatment $X$ can be conceptualised as having two binary components that act through different causal pathways: one component $X^c$ that affects the cancer of interest and one component $X^o$ that affects the competing event. 
The separable direct effect of treatment on the probability of death from cancer is defined as 
\begin{equation}
    E\left[F_c(t|X^c=1, X^o=x,\bm{Z})\right] -  E\left[F_c(t|X^c=0,X^o=x, \bm{Z})\right] 
\end{equation}
that is, the effect of the component of treatment that affects the event of interest when the component of treatment that affects the competing event $X^o$ is set to a constant value $x$, with $x=1$ or $x=0$.\\[2ex]

Analogously, we can define the separable indirect effect of treatment on the probability of death from cancer as 
\begin{equation}
    E\left[F_c(t|X^c=x, X^o=1,\bm{Z})\right] -  E\left[F_c(t|X^c=x,X^o=0, \bm{Z})\right] 
\end{equation}
that is, the effect of the component of treatment that affects the competing event when the component of treatment that affects the event of interest is set to a constant value.

The above definitions involve no hypothetical intervention of eliminating competing events as it is the case with direct effects.
However, separable effects assume a hypothetical intervention in which a different value is assigned in each component of the treatment.

\emph{Example}\\[2ex]

To estimate the separable effects for the prostate cancer example,  we need to make a copy of the treatment variable so that we can manipulate these separately in \texttt{standsurv}.
\begin{verbatim}
gen rx_c = rx
gen rx_o = rx
\end{verbatim}

We can now fit cause-specific models including either variable \texttt{rx\_c} or \texttt{rx\_o}:
\begin{verbatim}
// Prostate cancer
stset dtime, failure(eventType==1) exit(time 60)
stpm2 rx_c normalAct ageCat2 ageCat3 hx hgBinary, scale(hazard) df(4) ///
    tvc(rx_c) dftvc(2)
estimates store prostate

// Other causes
stset dtime, failure(eventType==2) exit(time 60)
stpm2 rx_o normalAct ageCat2 ageCat3 hx hgBinary, scale(hazard) df(3) 
estimates store other
\end{verbatim}

The parameters estimates are identical to the previous models and so are not shown.

Using a similar syntax as the one used to estimate the CIFs in section \ref{sec:total:crudeprob} and adding more \texttt{atn()} options we can get the separable direct and indirect effects:
\begin{verbatim}
standsurv, crmodels(prostate other) cif timevar(timevar) contrast(difference) ci ///
    at1(rx_c 1 rx_o 1)  ///
    at2(rx_c 1 rx_o 0)  ///
    at3(rx_c 0 rx_o 0)  ///
    atvars(F_rx11 F_rx10 F_rx00) contrastvars(F_diff_indirect F_diff_total)
\end{verbatim}

\begin{figure}[h!]
    \centering
    \includegraphics[width=0.9\textwidth]{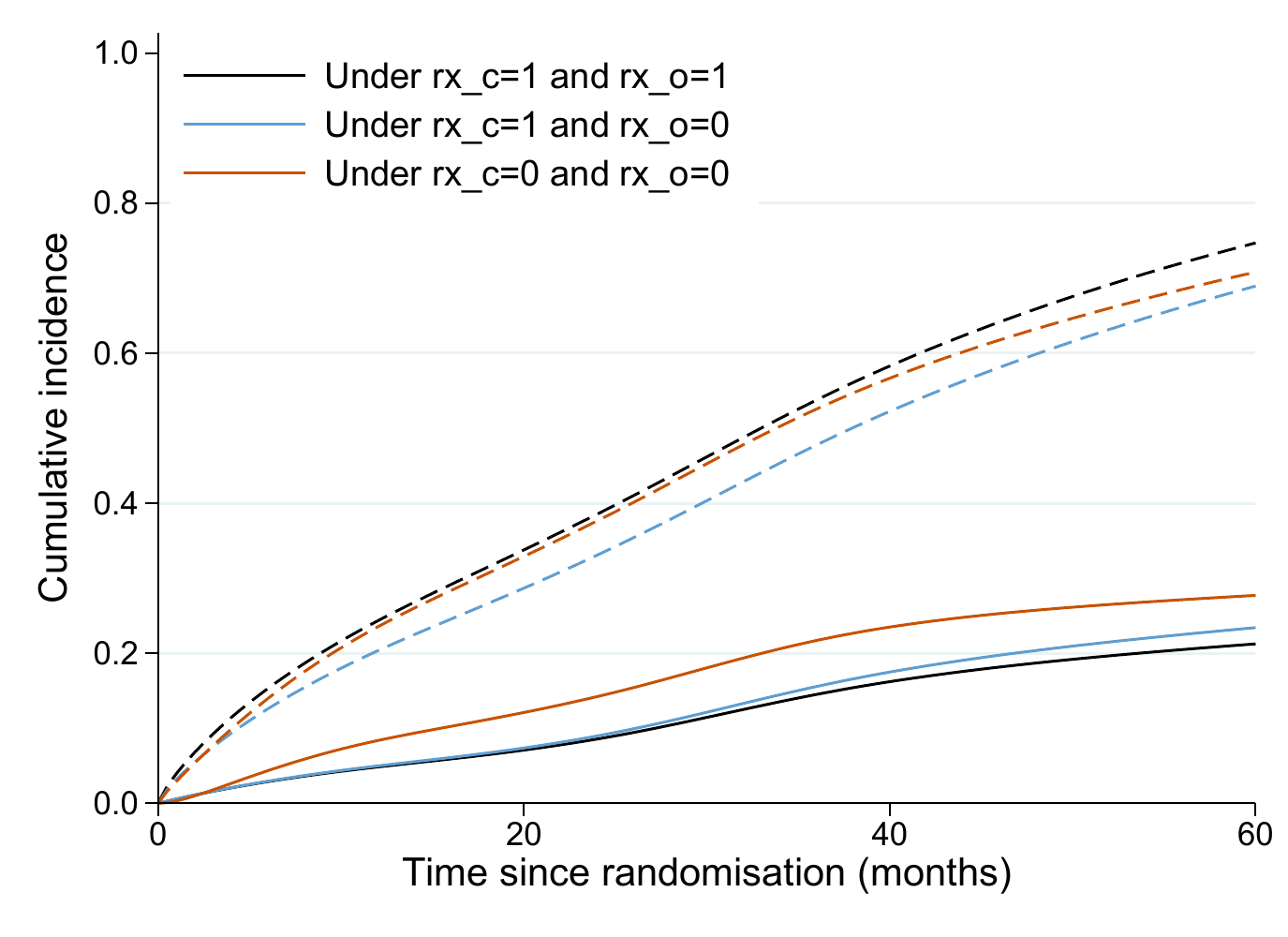}
    \caption{Cumulative incidence of death from prostate cancer (solid lines) and cumulative incidence of death from any cause (dash lines), under DES, under placebo as well as under the hypothetical treatment where the effect of other causes of death is removed (in blue).}
    \label{fig:totalsep}
\end{figure}

\begin{figure}
    \centering
    \includegraphics[width=0.9\textwidth]{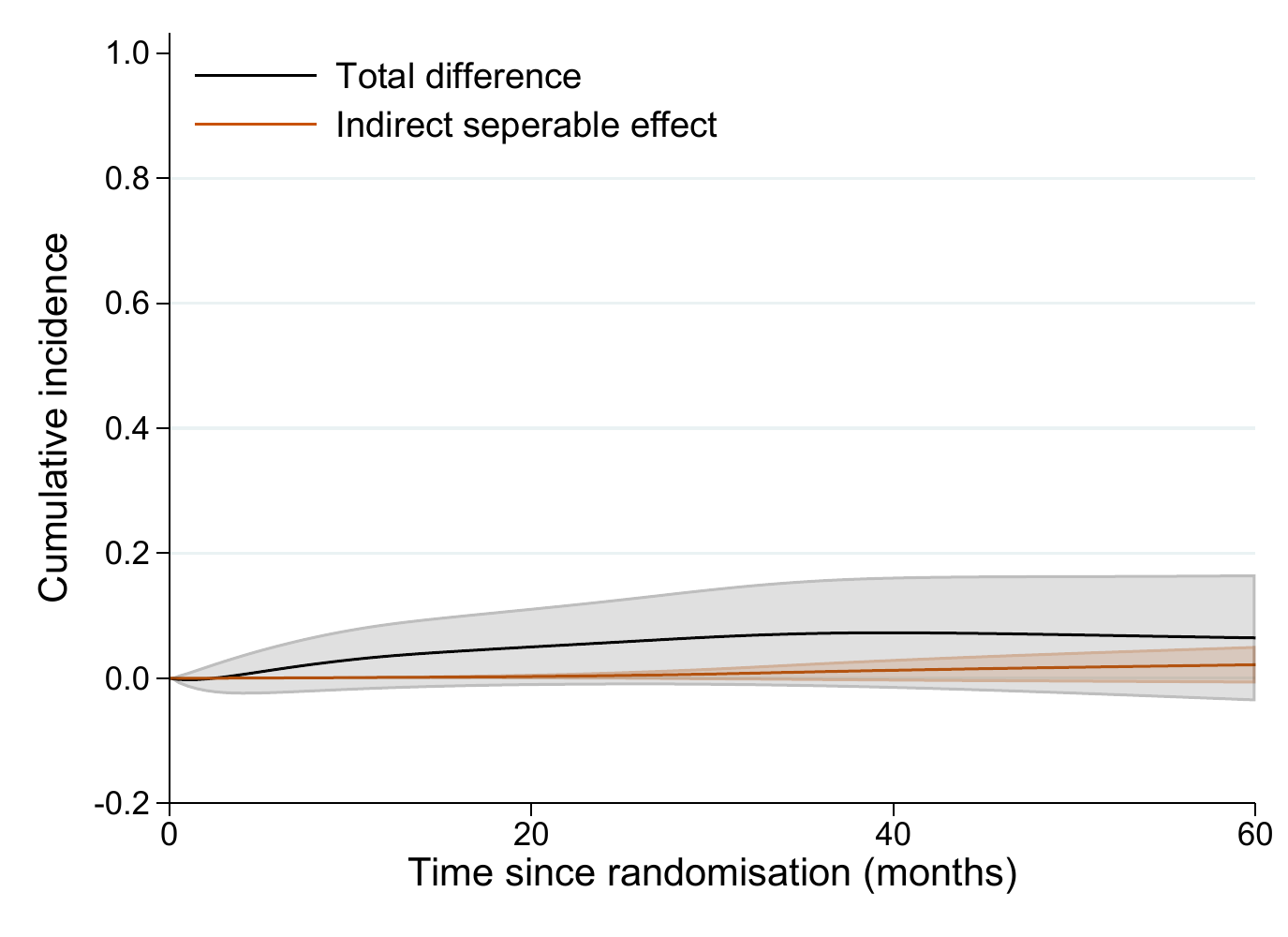}
    \caption{Standardised total difference in cumulative incidence of death from prostate cancer under DES and placebo by time since randomisation and the separable indirect difference.}
    \label{fig:sep}
\end{figure}

The standardised cumulative incidence of death from prostate cancer under DES (equal to 14.5\%, with 95\% CI: 9.8\%--21.5\%,  at 36 months since randomisation) and under placebo (equal to 21.7\%, with 95\% CI: 16\%--29.5\%, at 36 months since randomisation) as well as under the hypothetical treatment where the effect of other causes of death is removed (equal to 15.6\%, with 95\% CI: 10.6\%--23\%, at 36 months since randomisation) is shown in Figure \ref{fig:totalsep} (solid lines).
When the effect of other causes of death is removed, the cumulative incidence of prostate cancer death (blue line) is very close to the cumulative incidence of prostate cancer death under DES (black line), suggesting that the treatment effect is mainly driven by its effect on prostate cancer mortality.
The standardised total difference in the cumulative incidence of death from prostate cancer under DES and placebo as well as the separable indirect are given as a function of time since randomisation in Figure \ref{fig:sep}.
The indirect separable effect is increasing with time but remains low during the whole follow-up.
At 36 months (3 years) since diagnosis, when the total difference in standardised cumulative incidence of prostate death cancer is equal to 7.2\% (95\% CI:-1.4\%--15.8\%), the estimate of the indirect effect is 1.1\% (=15.6\%--14.5\%) with 95\% CI:-0.4\%--2.5\%. 
This corresponds to the reduction in prostate cancer mortality under DES compared to placebo that is due to the DES effect on mortality from other causes. 
Thus, the total effect of treatment on prostate cancer mortality is not highly driven by a harmful effect on death from other causes.

An interesting point here is that treatment has almost a null overall effect on the total probability of death (dash lines in Figure \ref{fig:totalsep}) and that this is due to the impact of treatment acting in opposite directions on the two competing causes of death. 
If we could imagine a treatment that only acted on the prostate mortality, but did not have the corresponding negative impact on other causes we can arrive at the blue line (i.e. a reduced deaths overall, and corresponding reduced deaths due to prostate cancer). 
This is the separable effect of treatment acting only on prostate cancer mortality.

\section{Discussion}\label{sec:discussion}
We have described causal effects that might be of interest in the presence of competing events and have shown how to estimate those using regression standardisation with the Stata command \texttt{standsurv}.
Identification assumptions for the causal effects described in this paper are discussed in detail elsewere \cite{Young2020, Stensrud2020}.
Causal effects can be defined as the total effect of treatment through all causal pathways between treatment and the event of interest (i.e. cumulative incidence and expected loss in life due to a cause of death) as well as the directs effect of treatment on the event of interest that does not capture the effect of treatment on the competing event (i.e. net probability of death). 
For settings where the  treatment effect can be decomposed into distinct components, separable effects have also been defined, with the separable indirect effect of treatment corresponding to the treatment effect on the event of interest only through its effect on the competing event.
We have demonstrated how to obtain estimates for all statistical estimands of interest and the causal effects with the post-estimation command \texttt{standsurv} using an example of publicly available prostate cancer data.
Even though the illustrative example is on cancer data the described methods are applicable also to other clinical areas.
Command \texttt{standsurv} applies regression standardisation and calculate the estimates as the average over all individual-specific predictions. 
Confidence intervals can also be derived using the delta method.

Total effects refer to a setting that entails no elimination of competing events while direct effects assume an intervention of eliminating competing events.
Each contrast has a different interpretation and the choice is based on the question of interest \cite{Young2020}.
An intervention of eliminating competing events might not be straightforward to realise in practice. 
For instance, it is not easy to think about an intervention that eliminates death.
Also, contrasts that are interpreted in a hypothetical world that is not possible to die from causes other than the event of interest are not useful for understanding the anticipated real‐world prognosis of patients.
For patients, clinicians, healthcare professionals and policymakers measures that refer to a setting where competing events are present is more relevant and total effects might be preferable.
However, the total effect of treatment on the event of interest has a challenging interpretation when treatment also affects the competing events; it provides no information about whether part of the treatment effect on the event of interest is due to the treatment effect on the competing event.
Reporting both total effects of treatment on the event of interest and competing events helps address this issue.
If interest is on comparing populations or across years with different background mortality rates, direct effects can be useful. 
Direct effects allow comparisons between populations without any possible distortions from competing causes of death.
They can also be applied to explore temporal trends or to study the aetiology of a disease. 
In general, using a variety of measures can help to understand different aspects of the impact of disease.

Separable effects can also be useful for situations where the effect of treatment can be partitioned into two components, one component affects the event of interest and one components affect the competing event through different causal pathways, and require no conceptual interventions on competing events (such as their elimination) \cite{Stensrud2020}.
When defining and interpreting separable effects, it is important to carefully consider a hypothetical intervention under which a different value is assigned in each component of the treatment so that there are well-defined effects. 
Sometimes decomposition of treatment might be difficult in practice circumventing verification of separable effects in a future experiment.
However, as others have argued, exploring a well-defined treatment decomposition within a formal causal framework can be a valuable tool for answering important research questions on whether treatment directly affects the event of interest, even if the decomposition is not possible in practice \cite{Stensrud2020}.

One of the total effects discussed in this paper was the expected loss in life due to a cause of death within a restricted time period.
The interpretation of this measure as life lost is more intuitive in comparison to other traditional measures such as probabilities.
However, it requires the choice of a pre-specified timepoint which add some complexity in its interpretation.
The expected loss in life makes also a comparison with an immortal cohort where patients are alive for the whole interval from 0 to time $t^*$ \cite{Mozumder2021}.
A measure with a more intuitive interpretation is the loss in expectation of life (LLE) or number of life years lost \cite{Andersson2013}. 
LLE compares the life-expectancy of patients to a comparable population group who are assumed to be disease-free and have similar characteristics and corresponds to the number of years that are lost due to the disease.
However, LLE requires extrapolation of the mortality rates beyond the available data.
To avoid strong extrapolation assumptions, the LLE within the first $t^*$ years (restricted LLE) could be estimated instead and this would provide a comparison of the disease population to the general population \cite{Royston2013}.

In cancer registry based studies, direct effects are usually referred to as the net setting and can be estimated using either the cause-specific approach or the relative survival approach.
The former approach was demonstrated in Section \ref{sec:direct}. 
However, the cause-specific approach requires appropriate classification of the cause of death.
As the cause of death information obtain by death certificates may not be available or not accurate, the relative survival approach is often preferred.
In the relative survival framework, separating deaths due to the cancer of interest from competing events (death due to other causes) is done indirectly by comparing all-cause survival in the cancer population to the survival of a comparable group of the general population with similar characteristics.
Causal effects in the relative survival framework can also be obtained using the \texttt{standsurv} command and these are discussed elsewhere \cite{Syriopoulou2020}.
A measure, conceptually similar to separable effects, has also been suggested in the relative survival framework; this is the avoidable deaths after an intervention that is assumed to affect only the cancer mortality rates and have no effect on the rates of other cause mortality \cite{Syriopoulou2021, Syriopoulou2020}.

In this paper, we have focused on baseline covariates that do not change over time.
Seperable effects also rely on the absence of time-varying covariates.
For settings where time-varying covariates may be present, various estimators has been suggested \cite{Young2020, Bekaert2010} and this also consists part of future work.
Finally, even though command \texttt{standsurv} was developed for obtaining marginal effects, it can also be used to obtain non-marginalised estimates.
This can be done by specifying the entire covariate pattern so that the predictions are not averaged over any covariate distribution and an example can be found in Appendix \ref{sec:advanceModelling}.

Several statistical estimands and causal effects can be defined in the presence of competing events and, under assumptions, estimates of those can be obtained using regression standardisation with the Stata command \texttt{standsurv}.
The choice of which causal effect to define should be given careful consideration based on the research question and the audience to which the findings will be communicated.

\subsection*{Abbreviations}
CI: confidence interval \\
CIF: cause-specific cumulative incidence functions \\
DES: diethylstilbestrol \\
FPM: flexible parametric survival model \\
RMFT: restricted mean failure time 

\subsection*{Funding}
PCL was supported by the Swedish Cancer Society (Cancerfonden) (Grant number 2018/744), the Swedish Research Council (Vetenskapsr\aa det) (Grant number 2017-01591) and Cancer Research UK (Grant number C1483/A18262). MJR was supported by a Cancer Research UK project grant (C41379/A27583). SIM was supported by the National Institute for Health Research (NIHR Advanced Fellowship, Dr Sarwar Mozumder, NIHR300100).

\subsection*{Availability of data and materials}
We use publicly available data from a trial on prostate cancer available at \url{https://hbiostat.org/data} \cite{Byar1980}.
Stata code for all the analysis is available at \url{https://github.com/syriop-elisa/competing_events_standsurv}.

\subsection*{Competing interests}
SIM works part-time (0.5 FTE) at Roche Products Limited not related to this research.

\appendix
\section{Data preparation} \label{sec:appendix}
We use data from a trial on prostate cancer (\texttt{prostate.dta}) to demonstrate how to obtain several measures of interest using regression standardisation with the Stata command \texttt{standsurv}. 
Data include 502 individuals that were randomly assigned estrogen therapy and are available at \url{https://hbiostat.org/data/} \cite{Byar1980}.
To prepare the data for the analysis we run the following commands
\begin{verbatim}
// load data
use "prostate", clear

// restrict to placebo and high dose estrogen 
keep if inlist(rx,1,4)

// update coding (0 for placebo, 1 for DES)
replace rx = cond(rx==1,0,1)
label define lblrx 0 "placebo" 1 "DES"
label values rx lblrx 

// replace follow-up time variable with half day if zero
replace dtime = 0.5 if dtime==0

// all cause indicator (i.e. death from any cause)  
gen allcause = status != 1
// event indicator (0: alive, 1: dead due to prostate cancer, 2: dead due to other causes)
gen eventType = cond(status==1,0,cond(status==2,1,2))
label define causelab 0 "Alive" 1 "Prostate" 2 "Other"
label values eventType causelab

//create categorical variables 
gen hgBinary = hg<12
egen ageCat = cut(age), at(0,60,75,100)
replace ageCat=1 if ageCat==60
replace ageCat=2 if ageCat==75
gen normalAct = pf == 1
\end{verbatim}

For the analysis, we use some user-written Stata commands.
These can be installed within Stata from the Boston College Statistical Software Components (SSC) archive as follows:
\begin{verbatim}
// To fit the flexible parametric survival models
 ssc install stpm2  
// To generate the restricted cubic spline functions
 ssc install rcsgen
\end{verbatim}

The \texttt{standsurv} command will be used to obtain marginal (and non-marginal) estimates using regression standardisation and it can be installed by running
\begin{verbatim}
net from https://www.pclambert.net/downloads/standsurv  
\end{verbatim}


\section{Advanced modelling details}\label{sec:advanceModelling}
For simplicity, in the main paper, we have only considered FPM with linear effects and no interactions between covariates.
However, these can easily be incorporated in the survival model.
Using \texttt{standsurv} we can, then, obtain estimates of interest in a similar way as in the previous sections but with further specifying the \texttt{atn()} options.
Below we provide some examples for obtaining cause-specific cumulative incidence after fitting more complex FPMs but other estimates of interest could also be obtained in a similar way.
For the remaining section, we keep the same model for other causes as the one described in Section \ref{sec:illustrativeExample} but allow more complex models for prostate cancer.
For instance, the interaction term for age and treatment can be generated by:
\begin{verbatim}
forvalues i = 2/3 {	 

    gen ageCat`i'rx=ageCat`i'*rx  		
    
}
\end{verbatim}

and included in the model:

\begin{verbatim}
stset dtime, failure(eventType==1) exit(time 60)

stpm2 rx normalAct ageCat2 ageCat3 hx hgBinary ageCat?rx, scale(hazard) df(4) ///
    tvc(rx) dftvc(2)

estimates store prostate
\end{verbatim}

Under this model, the marginal CIFs defined in Equations \eqref{crudeprobcancer} and \eqref{crudeother} can be estimated as the standardised CIFs by further specifying the \texttt{atn()} options for the interactions terms since these include the treatment of interest \texttt{rx}:
\begin{verbatim}
standsurv, crmodels(prostate other) cif timevar(timevar) contrast(difference) ci ///
    at1(rx 0 ageCat2rx 0 ageCat3rx 0)  ///
    at2(rx 1 ageCat2rx=ageCat2 ageCat3rx=ageCat3)  ///
    atvars(CIF0b CIF1b) contrastvars(CIF_diffb)
\end{verbatim}

We can also include non-linear effects in the survival model.
For example, instead of modelling age as a categorical variable, age can be modelled continuously allowing for non-linearity using restricted cubic splines.
To generate the restricted cubic spline functions in Stata the user-written command \texttt{rcsgen} can be used.

To generate restricted cubic splines with 4 knots (3 restricted cubic spline terms) for age at diagnosis:
\begin{verbatim}
rcsgen age, gen(agercs) df(3) orthog
// store knot positions in global macro
global ageknots `r(knots)'
// save matrix for orthogonalisation
matrix Rage =r(R)
\end{verbatim}
For 3 degrees of freedom, 3 new age spline variables are created, \texttt{agercs1} \-- \texttt{agercs3}. 
Here we store the knot locations and the ``R Matrix'', so that we can derive post-estimation predictions for specific ages later on.

\begin{figure}
    \centering
    \includegraphics[width=0.9\textwidth]{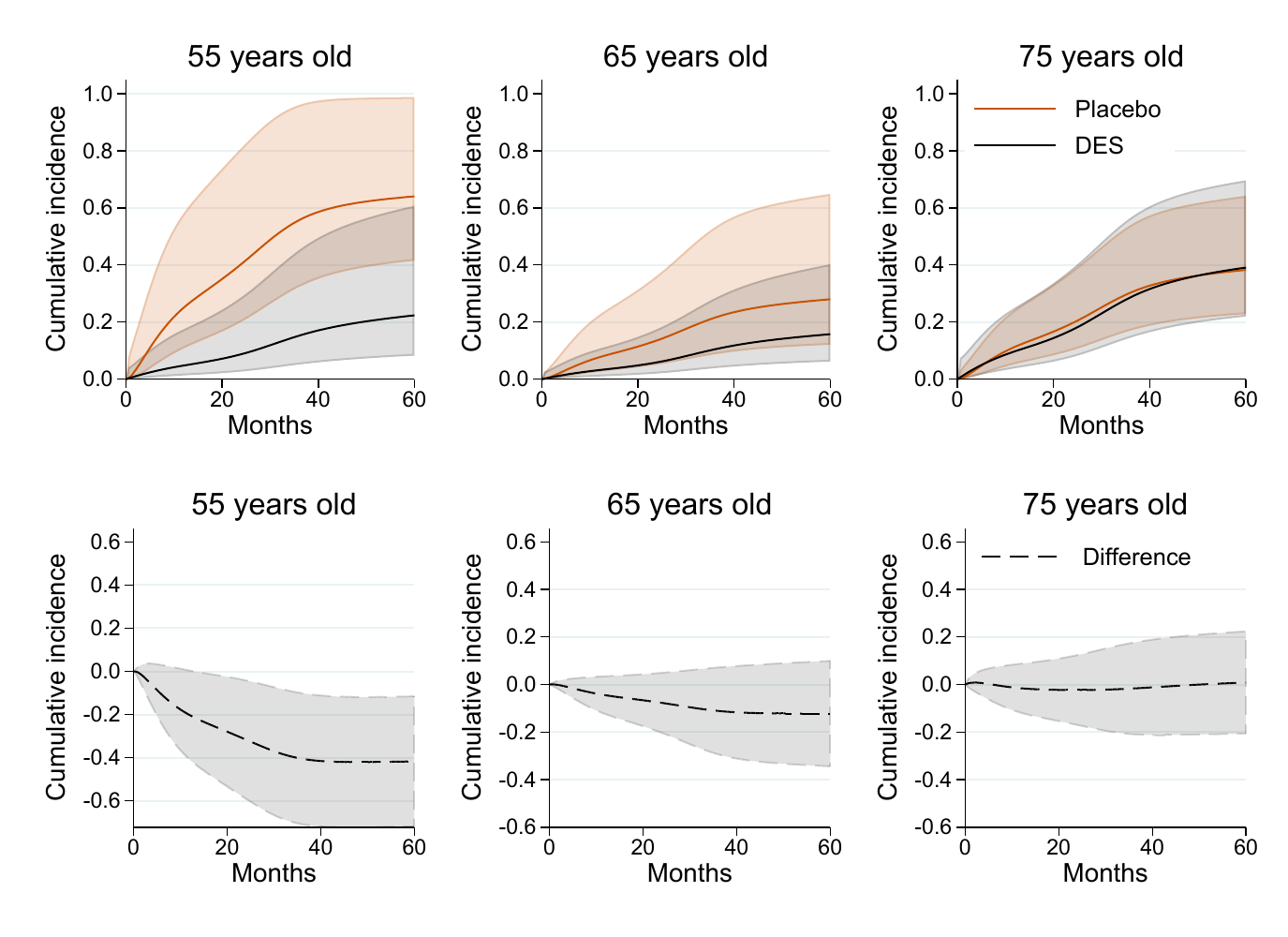}
    \caption{Age-specific cumulative incidence of death from prostate cancer under DES and placebo and their difference for individuals with normal daily activity, no history of cardiovascular disease and hemoglobin level lower than 12 (g/100ml) by time since randomisation, with 95\% confidence intervals.}
    \label{fig:total:agespec}
\end{figure}

Interactions involving the age splines can also be included in the model.
For instance, to generate interactions between age splines and treatment:
\begin{verbatim}
forvalues i = 1/3 {

    gen agercs`i'rx = agercs`i'*rx
    
}   
\end{verbatim}

The model can be fitted as:
\begin{verbatim}
stset dtime, failure(eventType==1) exit(time 60) 

stpm2 rx normalAct agercs1 agercs2 agercs3 hx hgBinary ///
    agercs1rx agercs2rx agercs3rx, ///
    scale(hazard) df(4) tvc(rx) dftvc(2) 

estimates store prostate
\end{verbatim}

To obtain the standardised CIFs under DES and under placebo from the above model: 
\begin{verbatim}
standsurv, crmodels(prostate other) cif timevar(timevar) contrast(difference) ci ///
    at1(rx 0 agercs1rx 0 agercs2rx 0 agercs3rx 0)  ///
    at2(rx 1 agercs1rx=agercs1 agercs2rx=agercs2 agercs3rx=agercs3)  ///
    atvars(CIF0c CIF1c) contrastvars(CIF_diffc)

\end{verbatim}

Even though command \texttt{standsurv} was developed for obtaining marginal effects, it can also be used to obtain non-marginalised estimates. 
These can be obtained by specifying the entire covariate pattern so that the predictions are not averaged over any covariate distribution. 
For instance, age-specific predictions can be derived by calculating the spline variables at that particular age with the same knot locations and projection matrix as before. 
An example is given below when interest is in the CIF of death from prostate cancer and we focus on individuals with normal daily activity (\texttt{normalAct}=1), no history of cardiovascular disease (\texttt{hx}=0) and hemoglobin level lower than 12 (g/100ml) (\texttt{hgBinary}=1) and compare CIFs of prostate cancer death under DES with CIFs under placebo, for ages 55, 65 and 75 years old.
Below, the spline variables for specific ages are stored in the local macros \texttt{c1}, \texttt{c2} and \texttt{c3}. 
\begin{verbatim}
foreach age in 55 65 75 {

    rcsgen, scalar(`age') knots($ageknots) rmatrix(Rage) gen(c)
    
    standsurv if _n==1, crmodels(prostate other) cif timevar(timevar) ///
        contrast(difference) ci ///
        at1(rx 0 normalAct 1 hx 0 hgBinary 1 ///
            agercs1 `=c1' agercs2 `=c2' agercs3 `=c3' ///
            agercs1rx 0 agercs2rx 0 agercs3rx 0) ///
        at2(rx 1 normalAct 1 hx 0 hgBinary 1 ///
            agercs1 `=c1' agercs2 `=c2' agercs3 `=c3' ///
            agercs1rx `=c1' agercs2rx `=c2' agercs3rx `=c3') ///
        contrastvars(CIF_diff`age') atvars(CIF0`age' CIF1`age') ///
        
}   
\end{verbatim}
As we do not average over each observation, we use \texttt{if \_n == 1} to tell \texttt{standsurv} to only take the first observation in the stacked data to calculate non-marginalised predictions. 
The age-specific CIF for individuals with normal daily activity, no history of cardiovascular disease and hemoglobin level lower than 12 (g/100ml) are shown in Figure \ref{fig:total:agespec}.
The difference in CIF of death from prostate cancer is large for young patients but the CIFs are almost identical for older ages.

\begin{figure}
    \centering
    \includegraphics[width=0.9\textwidth]{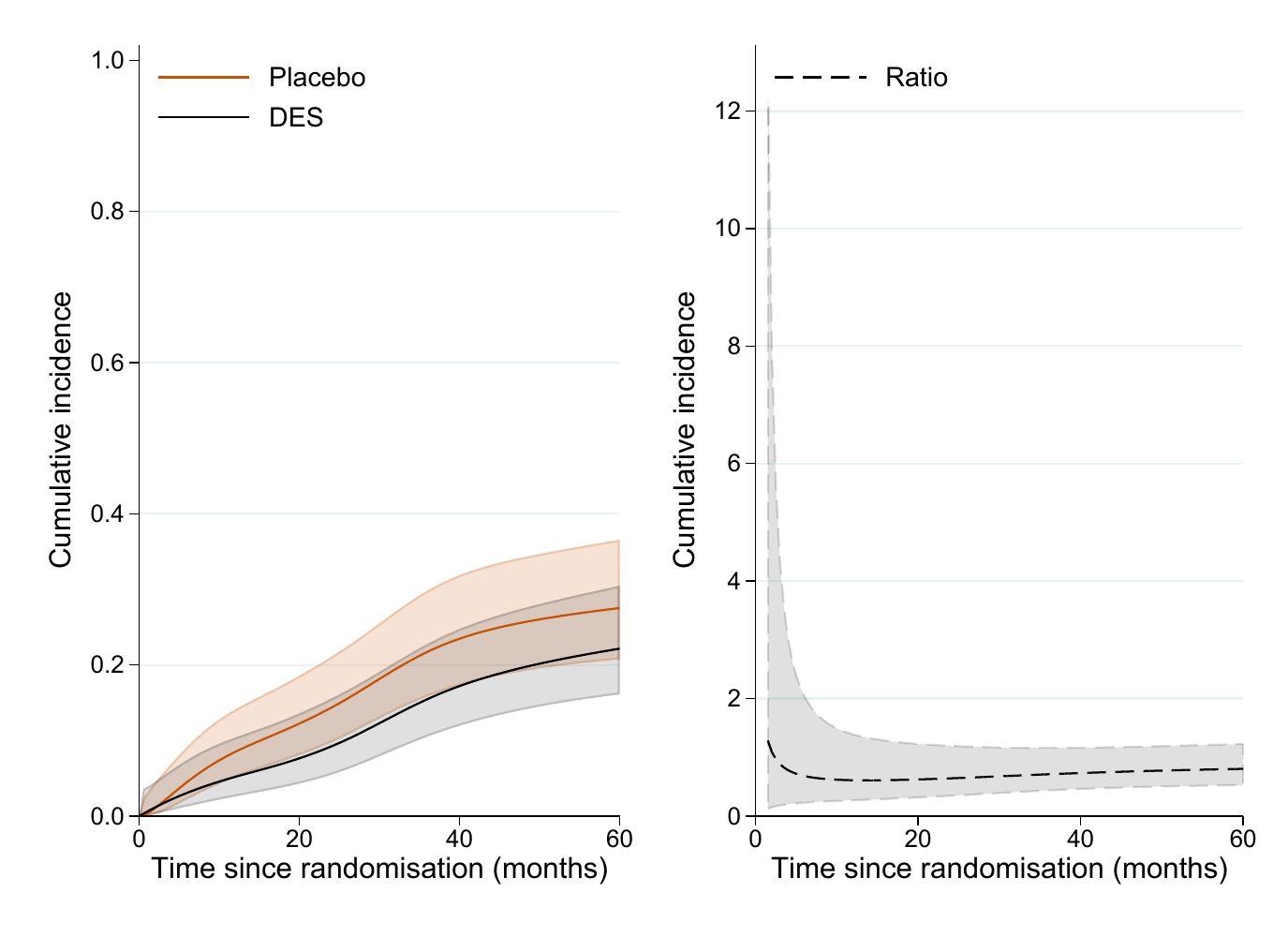}
    \caption{Standardised cumulative incidence prostate cancer under DES and placebo and their ratio (in black) by time since randomisation with 95\% confidence intervals.}
    \label{fig:total:ratio}
\end{figure}

Finally, in this paper the contrast of interest was defined as the difference under DES and placebo. 
Instead of the difference the ratio can also be calculated with the option \texttt{contrast(ratio)}. 
For instance, the ratio of standardised CIFs under DES and under placebo can be obtained by specifying \texttt{contrast(ratio)} within \texttt{standsurv} command:
\begin{verbatim}
standsurv, crmodels(prostate other) cif timevar(timevar) contrast(ratio) ci ///
    at1(rx 0 agercs1rx 0 agercs2rx 0 agercs3rx 0)  ///
    at2(rx 1 agercs1rx=agercs1 agercs2rx=agercs2 agercs3rx=agercs3)  ///
    atvars(CIF0d CIF1d) contrastvars(CIF_ratio)
\end{verbatim}
and this is shown in Figure \ref{fig:total:ratio} by time since randomisation.

In principle, any contrast can be obtained with \texttt{standsurv} by creating a user-defined mata function which can be called in the option \texttt{userfunction()} instead of the \texttt{contrast()}.

\bibliographystyle{unsrt}
\bibliography{bibliography}

\end{document}